\documentstyle[aps,prd,eqsecnum,twocolumn,epsf]{revtex}
\newcommand{\beq}{\begin{equation}}
\newcommand{\beqn}{\begin{eqnarray}} 
\newcommand{\eeq}{\end{equation}}
\newcommand{\eeqn}{\end{eqnarray}}
\newcommand{\beqa}{\begin{eqnarray}}
\newcommand{\eeqa}{\end{eqnarray}}

\newcommand{\gsim}{\mbox{\raisebox{-1.ex}{$\stackrel
     {\textstyle>}{\textstyle\sim}$}}}

\newcommand{\square}{\kern1pt\vbox{\hrule height
1.2pt\hbox{\vrule width 1.2pt\hskip 3pt
   \vbox{\vskip 6pt}\hskip 3pt\vrule width 0.6pt}\hrule
height 0.6pt}\kern1pt}

\def\beq{\begin{equation}}

\def\eeq{\end{equation}}

\begin{document}

\draft
\twocolumn[\hsize\textwidth\columnwidth\hsize\csname
@twocolumnfalse\endcsname

\title{{\bf Massive fermion production in nonsingular superstring 
cosmology\\}} 
\author{Shinji Tsujikawa and Hiroki Yajima} 
\address{Department of Physics, 
Waseda University, Shinjuku-ku, Tokyo 169-8555, Japan\\[.3em] 
e-mail:~shinji@gravity.phys.waseda.ac.jp, yajima@gravity.phys.waseda.ac.jp}
\date{\today} 
\maketitle
\begin{abstract}
We study massive spin-1/2 fermion production  in nonsingular superstring 
cosmology, taking into account one-loop quantum corrections to 
a superstring effective action with dilaton and modulus fields.  
While no creation occurs in the massless limit, massive fermions can be  
produced by the existence of a time-dependent frequency.  Due to the 
increase of the Hubble expansion rate during the modulus-driven phase, the 
occupation of number of fermions continues to grow until the point of the 
graceful exit, after which fermion creation ceases with the decrease of the 
Hubble rate.
\end{abstract}
%\pacs{PACS 98.80.Cq}
\vskip 2pc
]

%%%%%%%%%%%%%%%%%%%%%%%%%%%%%%%%%%%%%%
%                                                %
\section{Introduction}                           %
%                                                %
%%%%%%%%%%%%%%%%%%%%%%%%%%%%%%%%%%%

With the development of superstring theory, string-inspired 
cosmological models \cite{review} have received much attention 
to describe the evolution of the very early stage of the Universe.
Most of such scenarios are based on the low-energy effective
field theory, which is expected to be valid at the Planck scale.
While a full theory is not yet established, it is important to 
test the viability of string theories by extracting  various 
cosmological implications from them.

Among string-motivated cosmological models proposed so far,
the {\it pre-big-bang} (PBB) scenario \cite{PBB} has been 
most widely studied. 
If one assumes that the Universe has a $T$-duality, there 
exist two disconnected branches.  One of which ($t<0$) corresponds to the 
stage of superinflation driven by the kinetic term of the dilaton field, 
and another ($t>0$) is the Friedmann branch where the Universe exhibits  
standard decelerating expansion.  The PBB scenario basically requires the 
existence of nonsingular solutions which interpolates between two 
disconnected branches \cite{GV1}.  In the tree-level superstring action, 
however, one has no-go results that singularity can not be 
avoided \cite{tree,Kaloper}.

In order to overcome such singularity problems, 
Antoniadis, Rizos, and Tamvakis \cite{oneloop} involved one-loop quantum 
corrections to the string effective action with dilaton and modulus fields, 
and found some nonsingular solutions.
Since the success of singularity avoidance is mainly determined by the 
motion of the modulus field, the allowed ranges of 
parameters have been analyzed in the absence of the dilaton field in the flat 
Friedmann-Robertson-Walker (FRW) background \cite{RT} 
and the anisotropic Bianchi type-I metric \cite{KSS}.  
In the full dilaton-modulus system, 
several authors studied generality of singularity avoidance in the closed 
FRW \cite{closed} and Bianchi type-I and -IX \cite{yaji} spacetimes.  
It was found that nonsingular solutions generically exist 
except for the Bianchi-IX case.

From observational point of view, analysis of the perturbations predicted in  
the PBB model is an important issue in order to test realistic string theories.
In this respect, many authors investigated quantum creation of scalar 
particles such as dilaton and 
axions \cite{GVdilaton,CEW,BH2,DGSV,VMD,DKS}, 
and production of gravitational waves \cite{BGGV,BMU,CCG}, most of which 
exhibit different spectra compared to the standard cosmology.  Another 
interesting prediction in the PBB scenario is the generation of primordial 
magnetic fields due to the break of the conformal invariance 
\cite{gasperini,lemoine}.

Recently, Brustein and Hadad \cite{fermion} studied fermion production
in superstring cosmology in the presence of the dilaton coupling only.  It 
was found that massless fermions are not created, since the equation of 
fermions reduces to that in the Minkowski spacetime in the massless limit.  
In this {\it Letter}, we investigate {\it massive} fermion production in 
more general dilaton-modulus system whereby singularity can be avoided.
In fact we will show that massive fermions are nonadiabatically 
created due to the existence of a time-dependent mass term.

%%%%%%%%%%%%%%%%%%%%%%%%%%%%%%%%%%%%%%%%%%%%%%%%%%
 \section{Nonsingular solutions}
%%%%%%%%%%%%%%%%%%%%%%%%%%%%%%%%%%%%%%%%%%%%%%%%%%
Consider the following one-loop effective action of the heterotic superstring 
theory \cite{oneloop,closed,yaji},
\begin{eqnarray}
 S &=& \int d^4 x \sqrt{-g} \Biggl[ \frac12 R -\frac14(\nabla 
 \phi)^2-\frac34(\nabla \sigma)^2 \nonumber \\
 &-& \frac16 
 H_{\mu\nu\lambda}H^{\mu\nu\lambda}+\frac{1}{16} \left\{ \lambda 
 e^{\phi}-\delta \xi (\sigma) \right\} R_{\rm GB}^2 \Biggr],
\label{lag}
\end{eqnarray}
written in the Einstein frame. Here $R$, $\phi$, $\sigma$, and 
$H_{\mu\nu\lambda}$ are the scalar curvature, 
the dilaton, the modulus, and the antisymmetric 
tensor field, respectively.  In this work we set 
$H_{\mu\nu\lambda}=0$ and neglect the 
curvature terms higher than the second order.  The Gauss-Bonnet term, 
$R_{\rm GB}$, is defined as 
\begin{eqnarray}
R_{\rm GB}^2 =R^2-4R^{\mu\nu}R_{\mu\nu}+ 
R^{\mu\nu\alpha\beta}R_{\mu\nu\alpha\beta}.
\label{gauss}
\end{eqnarray}
In the presence of the last term in the action (\ref{lag}) (i.e, one-loop 
quantum corrections), singularity problems in the tree-level action can be 
avoided \cite{oneloop}.  The coefficients, $\lambda$ and $\delta$, are 
determined by the inverse string tension $\alpha'$ and the four-dimensional 
trace anomaly of the $N=2$ sector, respectively.  While $\lambda$ is 
positive definite, $\delta$ can be either positive or negative.

 The function, $\xi(\sigma)$, is expressed as 
\begin{eqnarray}
\xi(\sigma)={\rm ln} \left[2e^{\sigma-\pi e^{\sigma}/3}
\left\{ \prod_{n=1}^{\infty} \left( 1-e^{-2n\pi e^{\sigma}} 
\right) \right\}^4 \right].
\label{xi}
\end{eqnarray}
Then the first derivative of $\xi(\sigma)$ in terms of $\sigma$
is well approximated as $\xi'(\sigma) =-(2\pi/3) \sin {\rm h\sigma}$,
which we use in our numerical analysis.

It is also convenient to introduce a dimensionless function, $f(\phi,\sigma) 
\equiv [e^{\phi}-\bar{\delta} \xi(\sigma)]/16$ with 
$\bar{\delta}\equiv\delta/\lambda$.  We normalize time and spatial 
coordinates by the string length scale $\sqrt{\lambda}$ as 
$\bar{x^{\mu}}=x^{\mu}/\sqrt{\lambda}$, and scalar fields as 
$\bar{\phi}=\phi \sqrt{\lambda}$, $\bar{\sigma}=\sigma \sqrt{\lambda}$.  
Hereafter we drop bars for simplicity.

Adopting the flat FRW metric as the background spacetime, with $a \equiv 
e^p$ being the scale factor, the dynamical equations for the metric and 
scalar fields yield 
\beqa
& & 8(1+8\dot{p}\dot{f})(\ddot{p}+\dot{p}^{2})
+4(1+8\ddot{f})\dot{p}^2+\dot{\phi}^{2}
+3\dot{\sigma}^{2}=0,
\label{eq_p}\\
& & \ddot{\phi}+3\dot{p}\dot{\phi}
-2f_{,\phi} R^{2}_{\rm GB}=0, 
\label{eq_phi} \\
& & \ddot{\sigma}+3\dot{p}\dot{\sigma}
-\frac{2}{3}f_{,\sigma} R^{2}_{\rm GB}=0,
\label{eq_sigma}
\eeqa
together with the constraint equation,
\beqa
12\dot{p}^2+96\dot{p}^{3}\dot{f}
-\dot{\phi}^{2} -3\dot{\sigma}^{2}=0.
\label{constraint}
\eeqa
Here an overdot denotes a derivative with respect to cosmic time, $t$, 
and the Gauss-Bonnet term is given as $R^{2}_{\rm 
GB}=24\dot{p}^2(\ddot{p}+\dot{p}^{2})$.  Nonsingular cosmological 
solutions have been found for negative values of 
$\delta$ \cite{oneloop,closed}.

%%%%%%%%%%%%%%%%%%%%%%
\begin{figure}
\epsfxsize = 3.5in
\epsffile{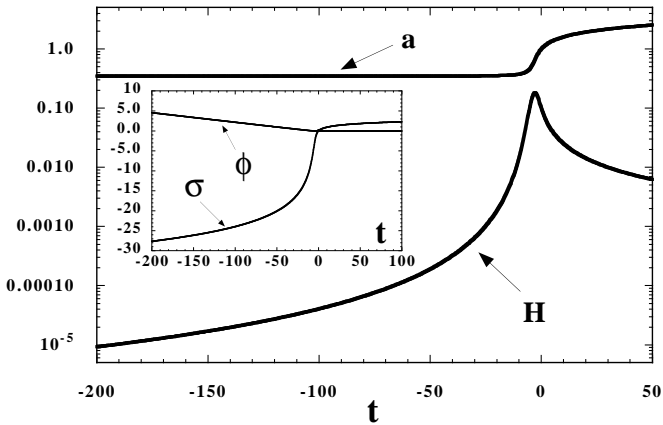} 
\caption{The evolution of the Hubble rate is plotted 
when the singularity is avoided.
We choose $\delta=-48/\pi$, and set  $\phi=\dot{\phi}=0$,
$\sigma=0$, $p=0$ at $t=0$.  $\dot{\sigma}$
is determined by the constraint equation (\ref{constraint})
as $\dot{\sigma}=0.2$.
{\bf Inset:} $\phi$ and $\sigma$ vs $t$.
The evolution of the system is dominated by the $\sigma$ 
field around the graceful exit.
}
\label{Fig1}
\end{figure}
%%%%%%%%%%%%%%%%%%%%%%

In the absence of the modulus field, the scale factor in the Einstein 
frame evolves as $a \propto |t|^{1/3}$ during the dilaton-driven phase.  
For negative $t$, this corresponds to the accelerated contraction, 
$\dot{a}<0$ and  $\ddot{a}<0$.  
In the tree-level action, one needs to assume 
that the epoch of the accelerated evolution comes to an end at some time in 
order to make a smooth transition to another branch ($t>0$).  In the 
present scenario, however, taking into account one-loop corrections opens 
up the possibility of the graceful exit driven by the kinetic energy of the 
modulus field.

We show one nonsingular solution in Fig.~1.  
Generally the Hubble rate, $H \equiv \dot{p}$, grows as 
$H \propto (-t)^{-2}$ during the modulus-driven phase.  In order to avoid 
singularity at $t=0$, the velocity of $\phi$ is required to be much smaller 
than $\sigma$ during the graceful exit. If $|\dot{\phi}|$ is sufficiently 
small around $t=0$, there always exist nonsingular solutions for $\phi<0$.  
This is because $e^\phi \approx 0$ for negative large values of $\phi$, 
which indicates that the singularity avoidance is practically independent 
of $\phi$.  The allowed parameter regions with respect to $\phi$ 
were precisely analyzed in Ref.~\cite{yaji}, which we do not 
repeat it here.

The sign of $\dot{\sigma}$ does not change during the whole evolution.
For $\dot{\sigma}>0$ which corresponds to the case of Fig.~1, $\sigma$ 
rapidly moves relative to $\phi$ around the graceful exit. 
After a smooth transition at $t=0$, the dilaton freezes with $\phi \sim 0$ and 
the modulus evolves as $\sigma \propto {\rm ln}~t$ as $t \to \infty$.  The 
Hubble rate begins to decrease for $t>0$, which asymptotically approaches 
the Friedmann-like Universe, $a \propto t^{1/3}$ and $H \sim 1/(3t)$.

%%%%%%%%%%%%%%%%%%%%%%%%%%
\section{Massive fermion production} 
%%%%%%%%%%%%%%%%%%%%%%%%%%

Let us consider the following action for the fermion field $\psi$
with bare mass $m$: 
\begin{eqnarray}
S_{\rm fermion} = \int d^4 x \sqrt{-g}~ 
f \left( i\bar{\psi}\bar{\gamma}^\mu \overrightarrow{D}_{\mu} \psi 
-m\bar{\psi} \psi \right),
\label{lagfermion}
\end{eqnarray}
where $\bar{\gamma}^\mu$ is the curved-space Dirac matrices, and  
$D_{\mu} \equiv \partial_{\mu}+(1/4)
\gamma_{\alpha \beta} \omega_{\mu}^{\alpha 
\beta}$ is the spin-1/2 covariant derivative, where 
$\omega_{\mu}^{\alpha \beta}$ is the spin connection.  
$\gamma_{\alpha}$
denotes the Dirac matrices in Minkowski spacetime with 
$\gamma_{\alpha \beta} \equiv 
\gamma_{[\alpha}\gamma_{\beta]}$.

{}From the action (\ref{lagfermion}) we obtain the Dirac equation, 
\begin{eqnarray}
\left( i \bar{\gamma}^\mu D_{\mu}-m+\frac{i}{2}
\frac{\partial_{\mu}f}{f} \bar{\gamma}^\mu
\right)\psi=0.
\label{dirac}
\end{eqnarray}
Since $\bar{\gamma}^0=\gamma^0$,
$\bar{\gamma}^i=\gamma^i/a$, and $\bar{\gamma}^\mu D_{\mu}
=\bar{\gamma}^\mu \partial_{\mu} +(3/2)H\gamma^0$
in the flat FRW background, Eq.~(\ref{dirac}) is simplified by 
introducing a new field, $\chi \equiv a^{3/2}f^{1/2}\psi$, as 
\begin{eqnarray}
\left( i \gamma^{\mu} \partial_{\mu}-ma \right)\chi=0,
\label{dirac2}
\end{eqnarray}
where $\partial_0$ denotes the derivative with respect to conformal 
time, $\eta \equiv \int a^{-1} dt$.  
We decompose the $\chi$ field into Fourier modes as 
\begin{eqnarray}
\chi \left( x \right) &=& \int \frac{d^3k}{(2\pi )^{3/2}} e^{-i\vec{k}\cdot
\vec{x}} \nonumber \\ 
&\times& \sum_{s=\pm 1} \left[ u_s(k,\eta)a_s(k) +
v_s(k,\eta)b_s^\dagger (-k)\right],
\label{decom}
\end{eqnarray}
where $v_s(k)=C{\bar u}^T_s (-k)$ with $C$ being a constant.

Defining 
$u_s=\left[u_+(\eta)\varphi_s(k), s u_-(\eta)\varphi_s(k) \right]^T$ and 
$v_s=\left[s v_+(\eta)\varphi_s(k), v_-(\eta)\varphi_s(k) \right]^T$ with 
$\varphi_s(k)$ being eigenvectors of helicity operators, the Dirac 
equation (\ref{dirac2}) reads \cite{GPRT,CKRT} 
\begin{eqnarray}
u'_{\pm}(\eta)=iku_{\mp}(\eta) \mp ima u_{\pm}(\eta),
\label{u_pm}
\end{eqnarray}
which reduces to the decoupled form: 
\begin{eqnarray}
u''_\pm+ \left[\omega_k^2 \pm i(ma)'\right]
u_\pm=0,
\label{decoupled}
\end{eqnarray}
where $\omega_k^2 \equiv k^2+(ma)^2$.  Note that we imposed the normalization 
conditions, $u_r^\dagger (k,\eta) v_s (k,\eta) =0$, $u_r^\dagger (k,\eta) 
u_s (k,\eta) = v_r^\dagger (k,\eta) v_s (k,\eta) =\delta_{rs}$, 
$|u_+|^2+|u_-|^2=2$.

In order to diagonalize the Hamiltonian,  we introduce  new operators, 
${\hat a}(k,\eta)=\alpha_k(\eta)a(k)+\beta_k(\eta) b^\dagger (-k)$ 
and ${\hat b}^\dagger (k,\eta)= -\beta^*_k(\eta)a(k)+ \alpha^*_k(\eta) 
b^\dagger (-k)$, where the Bogolyubov coefficients satisfy 
\begin{eqnarray}
\alpha_k=\frac{E_k+\omega_k}{F^*_k}\beta_k,~~|\beta_k|^2= 
\frac{|F_k|^2}{2\omega_k (\omega_k +E_k)},
\label{Bogolyubov}
\end{eqnarray}
with 
\begin{eqnarray}
&&E_k =k {\rm Re} (u_+^*u_-) +ma
\left( 1-|u_+|^2 \right) , \\
&&F_k = (k/2) \left(u_+^2 -u_-^2\right) + ma
u_+u_- .
\label{EKFK}
\end{eqnarray}
Note that the canonical commutation relation leads to 
$|\alpha_k|^2+|\beta_k|^2=1$, which restricts the occupation 
numbers of fermions as $n_k \equiv |\beta_k|^2 \le 1$. 
The initial conditions are chosen as $u_{\pm}(\eta_0)=
 \sqrt{(\omega_k \mp ma)/\omega_k}$, 
 corresponding to $E_k(\eta_0)=\omega_k$, $F_k(\eta_0)=0$, 
and $n_k(\eta_0)=0$.

%%%%%%%%%%%%%%%%%%%%%%
\begin{figure}
\epsfxsize = 3.5in
\epsffile{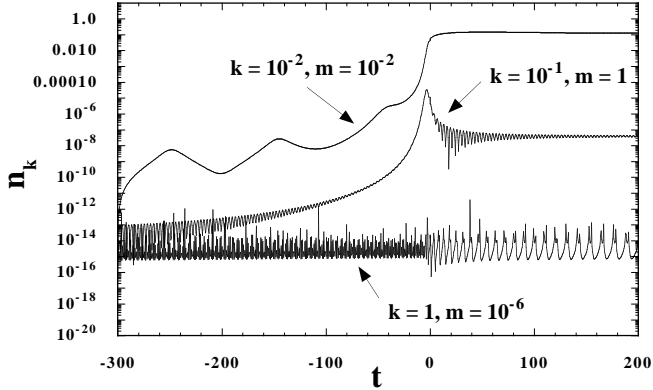} 
\caption{The evolution of the occupation number of 
fermions for three cases:
$k=10^{-2}, m=10^{-2}$; $k=10^{-1}, m=1$; and
$k=1, m=10^{-6}$,
where $k$ and $m$ are normalized by the string length scale,
$\sqrt{\lambda}$.
The $\delta$ and initial values of background quantities
are the same as in Fig.~1.  The enhancement of fermions 
strongly depends on the relation of two parameters, 
$k$ and $m$.}
\label{Fig2}
\end{figure}
%%%%%%%%%%%%%%%%%%%%%%

In the limit of  $m \to 0$, Eq.~(\ref{decoupled}) reduces to that in Minkowski 
spacetime, 
\begin{eqnarray}
u''_\pm+ k^2u_\pm=0.
\label{Minkowski}
\end{eqnarray}
The solution for this equation is expressed as 
$u_{\pm}(\eta)=e^{ik(\eta-\eta_0)}$, where we used the initial conditions, 
$u_{\pm}(\eta_0)=1$.  Then we have $n_k=0$ by 
Eqs.~(\ref{Bogolyubov})-(\ref{EKFK}), which indicates that no creation 
occurs in the massless limit \cite{fermion}.

When the mass of fermion is sufficiently small relative to the 
physical wave number ($m \ll k/a$), the situation is similar to 
the massless case.  In Fig.~2 we find that fermions are hardly excited for 
$k=1$ and $m=10^{-6}$, where $k$ and $m$ are normalized by 
the string length scale, $\sqrt{\lambda}$.
Note that the acquired number of e-foldings during 
the modulus-driven phase is not large compared to the standard inflationary
scenarios, e.g., if we normalize the scale factor as $a=1$ for $t=0$, $a 
\approx 0.345$ for $t=-500$ at which the contribution of modulus begins 
to be important relative to dilaton.  This indicates that when 
the condition, $m \ll k/a$, holds in the initial stage of the 
modulus-driven phase, it is typically valid even around the graceful exit.

%%%%%%%%%%%%%%%%%%%%%%
\begin{figure}
\epsfxsize = 3.5in
\epsffile{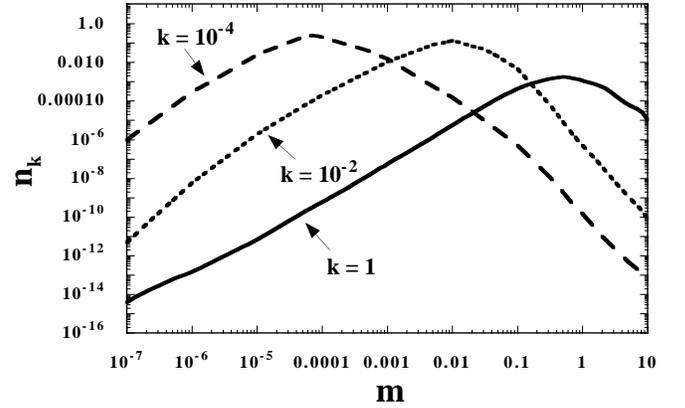} 
\caption{The occupation number of fermions
at the end of the modulus-driven phase 
as a function of the fermion mass, $m$, for three different
comoving momenta, $k=10^{-4}$, $k=10^{-2}$, and $k=1$.
The $\delta$ and initial values of background quantities
are the same as in Fig.~1.
}
\label{Fig3}
\end{figure}
%%%%%%%%%%%%%%%%%%%%%%

When $m$ and $k/a$ are comparable during the modulus-driven phase,
massive fermions are created due to the existence
of the time-dependent mass term in Eq.~(\ref{decoupled}).
In Fig.~2 the number density of fermions continues to grow until the point of 
the graceful exit for the cases of $k=10^{-2}, m=10^{-2}$; and $k=10^{-1}, 
m=1$.  The nonadiabatic condition where particles are sufficiently excited
can be written as $|\dot{\omega}_k|~\gsim~\omega_k^2$, which yields 
\begin{eqnarray}
H~\gsim~\frac{\left[k^2+(ma)^2\right]^{3/2}}{(ma)^2}.
\label{nonadiabatic}
\end{eqnarray}
In a dust or radiation dominated Universe, the Hubble rate decreases as 
$H \propto 1/t$, which works to violate the nonadiabatic condition, 
(\ref{nonadiabatic}).  In fact, in preheating after inflation, unless an
inflaton decay to fermions is not taken into account, 
the mass term $(ma)^2$ does not lead to sufficient fermion production 
\cite{GPRT,BHP,GK,shinji}.  
However, in the present model, the growth of the Hubble rate during the 
modulus-driven phase assists the nonadiabatic condition to hold, which 
results in nonperturbative particle creation solely by the time-dependent mass 
term.  The growth of the occupation number ends after the smooth transition 
to the Friedmann-like Universe, since the Hubble rate begins to decrease 
(see Fig.~1).

When $m$ is much larger than $k/a$, fermion production is generally 
suppressed.  Especially for $m \gg H$ where the nonadiabatic condition
(\ref{nonadiabatic}) is not satisfied, Eq.~(\ref{decoupled}) is approximately 
written as $\ddot{X}_{\pm}+m^2 X_{\pm} \simeq 0$ 
with $X_{\pm} \equiv a^{1/2}u_{\pm}$.  
This indicates that fermions are 
hardly created in the supermassive limit, $m \to \infty$.

In Fig.~3 we plot the occupation number of 
fermions at $t=0$ as a function of mass for three different momenta.
For each momentum, there exists a maximum $n_k$
for some value of $m$ ($=m_*$).
Since $m_*$ is typically of the same order as the each corresponding 
momentum, the curves shift from left to right with increasing $k$.
If the $1/\sqrt{\lambda}$ is  around the Planck scale, our results suggest that 
massive fermions heavier than the GUT scale can be copiously produced, 
which may play important roles 
for the leptogenesis scenarios \cite{GPRT}.

%%%%%%%%%%%%%%%%%%%%%%%%%%%%%%%%%
 \section{Conclusions and discussions}
%%%%%%%%%%%%%%%%%%%%%%%%%%%%%%%%%

We have investigated the production of massive spin-1/2 fermions 
in nonsingular 
superstring cosmology with dilaton and modulus fields.  The existence of 
the modulus coupled to the Gauss-Bonnet curvature invariant leads to a smooth 
transition from the modulus-driven accelerated expansion phase toward the 
Friedmann-like Universe.  Since the Hubble rate increases before the 
graceful exit, this makes it possible to produce massive fermions in terms of 
nonadiabatic change of their frequencies.  In particular, fermions are most 
efficiently excited when the bare mass $m$ is comparable to the physical 
momentum, $k/a$.  In both massless and supermassive limits, 
fermion creation is strongly suppressed.

The occupation number of fermions achieved 
by the time-dependent mass term $(ma)^2$ is typically smaller 
than the Pauli bound, $n_k=1$, even at the end of 
the modulus-driven phase.  If one introduces the Yukawa couplings 
between two scalar fields $\phi, \sigma$ and the fermion $\psi$ 
such as $h_1\phi \bar{\psi}\psi$ and $h_2\sigma \bar{\psi}\psi$, 
the effective mass of fermions is expressed as \cite{shinji} 
\begin{eqnarray}
m_{\rm eff}=m+h_1\phi+h_2\sigma.
\label{effmass}
\end{eqnarray}
In this case it is known that particle creation is most efficient when 
$m_{\rm eff}$ vanishes, leading to $n_k \sim 1$ 
both in the context of inflation 
\cite{CKRT} and preheating \cite{GPRT,BHP,GK,shinji}.  In the present 
scenario, since $m_{\rm eff}$ vanishes or becomes close to zero depending 
on two coupling constants, $h_1$ and $h_2$, this may further strengthen 
nonadiabatic amplification of fermions.  We leave to future work about the 
precise investigation of this issue.

Although we have restricted ourselves in spin-1/2 fermions satisfying the 
Dirac equation, nonthermal production of gravitinos (spin-3/2 fermions) has 
recently become an issue of great importance \cite{gravitinos}.  Gravitinos 
have both helicity-3/2 and -1/2 states.  While the helicity-3/2 mode 
reduces to the form of the Dirac equation, the helicity-1/2 mode behaves 
like the goldstino in global supersymmetric limit.  It was shown in 
Ref.~\cite{fermion} that the latter mode also reduces to the same form as 
the former mode in the massless limit by assuming the power-law evolution 
of the scale factor in the standard PBB scenario, which means that creation 
of massless gravitinos is highly suppressed.  However, the situation will 
change for {\it massive} gravitinos, in which case the existence of 
time-dependent mass terms may lead to the amplification of gravitinos 
during the modulus-driven phase.  It is certainly of interest to study 
gravitino production in realistic nonsingular cosmology, since this 
provides a powerful mechanism to distinguish between different models of 
string theories and regions of the parameter space, together with the CMB 
constraint by scalar and tensor metric perturbations produced during the 
graceful exit \cite{KS2}.

%%%%%%%%%%%%%%%%%%%%%%%%%%%%%%%%%%%%%%%%%%%%%%%
\section*{ACKOWLEDGMENTS}
We thank Bruce A. Bassett and Kei-ichi Maeda for useful discussions.  This 
work was supported by the Waseda University Grant for Special Research 
Projects.  
%%%%%%%%%%%%%%%%%%%%%%%%%%%%%%%%%%%%%%%

% references

\end{document}